\pdfoutput=1
\RequirePackage{ifpdf}
\ifpdf % We are running pdfTeX in pdf mode
\documentclass[pdftex]{sigma}
\else
\documentclass{sigma}
\fi

\def\diag{\operatorname{diag}}

\def\vphi{\varphi}

\begin{document}
\allowdisplaybreaks

\renewcommand{\thefootnote}{$\star$}

\renewcommand{\PaperNumber}{105}

\FirstPageHeading

\ShortArticleName{Separation of Variables and Contractions on Two-Dimensional Hyperboloid}

\ArticleName{Separation of Variables and Contractions\\ on Two-Dimensional Hyperboloid\footnote{This
paper is a~contribution to the Special Issue ``Superintegrability, Exact Solvability, and Special Functions''.
The full collection is available at
\href{http://www.emis.de/journals/SIGMA/SESSF2012.html}{http://www.emis.de/journals/SIGMA/SESSF2012.html}}}

\Author{Ernie KALNINS~$^\dag$, George S.
POGOSYAN~$^{\ddag\S}$ and Alexander YAKHNO~$^{\S}$}

\AuthorNameForHeading{E.~Kalnins, G.S.~Pogosyan and A.~Yakhno}

\Address{$^\dag$~Department of Mathematics, University of Waikato, Hamilton, New Zealand}
\EmailD{\href{mailto:math0236@math.waikato.ac.nz}{math0236@math.waikato.ac.nz}}

\Address{$^{\ddag}$~International Center for Advanced Studies, Yerevan State University,\\
\hphantom{$^{\ddag}$}~A.~Manougian 1, Yerevan, 0025, Armenia}
\EmailD{\href{mailto:pogosyan@theor.jinr.ru}{pogosyan@theor.jinr.ru}}

\Address{$^{\S}$~Departamento de Matem\'{a}ticas, CUCEI, Universidad de Guadalajara, Jalisco,
Mexico}
\EmailD{\href{mailto:alexander.yakhno@cucei.udg.mx}{alexander.yakhno@cucei.udg.mx}}

\ArticleDates{Received August 09, 2012, in f\/inal form December 19, 2012; Published online
December 26, 2012}

\Abstract{In this paper analytic contractions have been established in the $R\to\infty$
contraction limit for exactly solvable basis functions of the Helmholtz equation on the
two-dimensional two-sheeted hyperboloid.
As a~consequence we present some new asymptotic formulae.}


\Classification{70H06; 35J05}

\renewcommand{\thefootnote}{\arabic{footnote}}
\setcounter{footnote}{0}

\section{Introduction}

In the recent series of papers~\cite{IPSW1,IPSW3,IPSW2,KMP1,PSW1,PW1,
YAP1,YAP2,YAP3} a~special class of In\"onu--Wigner contractions was
introduced, namely ``analytic contractions''.
For an analytic
contraction the contraction parameter $R$, which is the radius of an
$n$-dimensional sphere or pseudosphere
$u_0^2+\varepsilon{\vec u}^2=R^2$, $(\varepsilon=\pm1)$
f\/igures in the separated coordinate systems, and in the eigenfunctions
and the eigenvalues for the Laplace--Beltrami operator (or Helmholtz equation).
With the help of analytic contractions we have established the connection
between the procedure of separation of variables for homogeneous spaces with
constant (positive or negative) curvature and f\/lat spaces.
For instance
it has been indicated how the systems of coordinates on the sphere~$S_2$
and hyperboloid~$H_2$ transform to four systems of coordinates on Euclidean
space~$E_2$.

The goal of this note is to establish the contraction limit $R\to\infty$
for eigenfunctions (or basis functions) of the two-dimensional
Helmholtz equation on the two-sheeted hyperboloid
$H_2:u_0^2-u_1^2-u_2^2=R^2$,
\begin{gather}\label{HE1}
\Delta_{\rm LB}\Psi=-\frac{\sigma(\sigma+1)}{R^2}
\Psi,
\qquad
\sigma=-1/2+i\rho,
\end{gather}
where the Laplace--Beltrami operator in the curvilinear
coordinates $(\xi^1,\xi^2)$ has the form
\begin{gather*}%\label{ALGEBRA4}
\Delta_{\rm LB}=\frac{1}{\sqrt{g}}\frac{\partial}{\partial\xi^i}
\sqrt{g}g^{ik}\frac{\partial}{\partial\xi^k},
\qquad
g=|\det(g_{ik})|,
\qquad
g_{ik}g^{k\mu}=\delta^\mu_i,
\end{gather*}
with the following relation between local metric tensor $g_{ik}(\xi)$,
($i,k=1,2$) and the ambient space metric $G_{\mu\nu}=\diag(1,-1,-1)$,
\begin{gather*}%\label{METRIC}
g_{ik}(\xi)=G_{\mu\nu}\frac{\partial u^\mu}{\partial\xi^i}
\frac{\partial u^\nu}{\partial\xi^k},
\qquad
i,k=1,2,
\qquad
\mu,\nu=0,1,2.
\end{gather*}
It is well known that the Helmholtz equation~\eqref{HE1} on the two-sheeted
hyperboloid admits separation of variables in nine orthogonal systems
of coordinates~\cite{OLEV,4}.
These nine systems of coordinates can be separated into three classes.
The f\/irst class includes three systems of coordinates which are of subgroup type: viz.\
pseudo-spherical, equidistant and horicyclic coordinates, the second class
includes three non subgroup type of coordinates: semi circular parabolic,
elliptic parabolic and hyperbolic parabolic.
The last two coordinate systems
in the general case contain the dimensionless parameter $\gamma$ and in
the limiting cases when $\gamma\to0$ or $\gamma\to\infty$ transform to
horicyclic or equidistant systems of coordinates.
These systems of coordinates
have an important property; the Helmholtz equation~\eqref{HE1} admits
the exact solution (which we call exactly solvable) in terms
of ``classical" special functions, namely Bessel and Legendre functions
or hypergeometric one.
The third class of coordinates consists of elliptic, hyperbolic and semi hyperbolic
coordinates each of which belong to the class of non subgroup type.
Two of them,
elliptic and hyperbolic coordinates, also contain a~dimensionless
parameter which is included in the metric tensor and Laplace--Beltrami operator.
The Helmholtz equation~\eqref{HE1} after the separation of variables in these
coordinates leads to the two Heun type dif\/ferential equations with a~four
singular points (which we call non-exactly solvable equations) and whose
solutions appear as Lame or Lame--Wangerin functions~\cite{BE3, GROSCHE, 2}.
In contrast to the classical special
functions which can be appear in terms of hypergeometric functions, the
Lame or Lame--Wangerin functions are def\/ined in terms of inf\/inite series
where the expansion coef\/f\/icients can not be written in explicit form because
they are the subject the third or higher order recurrence relations.

The contraction limit of a~basis function is not a~trivial task in the
case of exactly or non-exactly solvable equations.
Some calculations come
from the papers~\cite{IPSW1,IPSW3,IPSW2}, but many of them are not known
till now.

In this paper we restrict ourselves to the contraction limit $R\to\infty$
for four kinds of orthogonal basis functions: horocyclic, semi circular
parabolic, elliptic parabolic and hyperbolic parabolic, which is new.
We shall present normalizable eigenfunctions but do not give their
normalization constants explicitly, except in the case of horocyclic coordinates
which is particularly simple.
Here we have also included for completeness
the contraction for pseudo spherical and equidistant wave functions
(see~\cite{IPSW2,PSW1}).

We hope that our results beside possible application to the theory of special
functions will be useful also in the investigation of super integrable systems
which admit separation of variables on the two-dimensional hyperboloid.
We think
that they can be generalized for the three and higher dimensional hyperbolic
space when these six systems of coordinates are the sub systems of more complicated
systems of coordinates and where the procedure of variable separation leads to
similar dif\/ferential equations.

\section{Contractions}

\subsection{Pseudo-spherical basis to polar}

The f\/irst coordinate system is the pseudo-spherical system
($\tau>0$, $\varphi\in[0,2\pi)$)
\begin{gather*}%\label{spherical}
u_0=R\cosh\tau,\qquad
u_1=R\sinh\tau\cos\vphi,\qquad
u_2=R\sinh\tau\sin\vphi,
\end{gather*}
and the orthogonal basis functions of equation~\eqref{HE1} are
\cite{IPSW2,2,PSW1,PW1}{\samepage
\begin{gather*}%\label{sol_S}
\Psi^{\rm S}_{\rho m}(\tau,\vphi)=P^{|m|}_{i\rho-1/2}(\cosh\tau)
e^{i m\vphi},
\end{gather*}
where $P^{\mu}_{\nu}(z)$ is the Legendre function.}

In the contraction limit $R\rightarrow\infty$ we have
\begin{gather*}
\sinh\tau \sim \tau \sim \frac{r}{R}, \qquad
\vphi \to \vphi,
\qquad
\rho \sim k R,
\end{gather*}
where $(r,\vphi)$ are the polar coordinates in Euclidean plane.
There is now the matter of how this contraction af\/fects the basic
eigenfunctions that can be computed on the hyperboloid.
Using the well-known representations of the Legendre function in terms of
the hypergeometric function~\cite{BE1},
\begin{gather*}
P^{|m|}_{i\rho-1/2}(\cosh\tau) =
\frac{\Gamma(1/2 + i\rho + |m|)}{\Gamma(1/2 + i\rho - |m|)}
\left(\sinh{\frac{\tau}{2}}\right)^{|m|}
\left(\cosh{\frac{\tau}{2}}\right)^{|m|}
\frac{1}{|m|!}
\\
\hphantom{P^{|m|}_{i\rho-1/2}(\cosh\tau) =}{}\times
{}_2F_1 \left(1/2 + i\rho +|m|, 1/2 - i\rho +|m|; 1+|m|;
-\sinh^2{\frac{\tau}{2}}\right),
\end{gather*}
%and %using
asymptotic formulae for gamma-functions at large $z$~\cite{BE1},
\begin{gather*}
\frac{\Gamma(z+\alpha)}{\Gamma(z+\beta)}\approx z^{\alpha-\beta},
\end{gather*}
and taking into account that at $R\to\infty$
\begin{gather*}
F\left(\frac{1}{2}+|m|+i\rho,
\frac{1}{2}+|m|-i\rho;1+|m|;-\sinh^2\frac{\tau}{2}\right)
\\
\qquad\approx{_0F_1}\left(1+|m|;-\frac{k^2r^2}{4}\right)=\left(\frac{2}{kr}\right)^{|m|}|m|!J_{|m|}(kr),
\end{gather*}
where $J_{\nu}(z)$ is the Bessel function~\cite{BE2}, we get
\begin{gather*}
P^{|m|}_{i\rho-1/2}(\cosh\tau) \approx P^{|m|}_{ikR-1/2}
\left(1+ \frac{r^2}{2R^2}\right) \approx (-kR)^{|m|} J_{|m|}(kr).
\end{gather*}
Finally, the pseudo-spherical functions in the contraction limit
$R\rightarrow\infty$ take the form
\begin{gather*}
\Psi_{\rho m}^{\rm S}(\tau,\varphi)\approx(-kR)^{|m|}J_{|m|}(kr)
e^{im\varphi},
\end{gather*}
i.e., the pseudo-spherical basis up to the constant factor contracts
into polar one.
The correct correspondence to give limiting orthogonality relations
in the polar coordinates $r$ and $\vphi$ can be obtained using the contraction
limit of the normalization constant (see~\cite{IPSW2}) and the results
\begin{gather*}
\sinh\tau d\tau d\vphi \to \frac{1}{R^2} r d r d\vphi,
\qquad
\delta(\rho - \rho') \to \frac{1}{R} \delta (k-k'),
\end{gather*}
and
\begin{gather*}
\int_0^{\infty} J_{|m|}(kr) J_{|m|}(k'r) r dr = \frac{1}{k} \delta (k-k').
\end{gather*}

\subsection{Equidistant basis to Cartesian}

The coordinate system is the following one ($\tau_1,\tau_2\in\mathbb{R}$) %:     one?
\begin{gather*}%\label{sys_equi}
u_0=R\cosh\tau_1\cosh\tau_2,\qquad
u_1=R\cosh\tau_1\sinh\tau_2,\qquad
u_2=R\sinh\tau_1,
\end{gather*}
From the above def\/inition we have that
\begin{gather*}%\label{coor1}
\sinh\tau_1=\frac{u_2}{R},
\qquad
\tanh\tau_2=\frac{u_1}{u_0}
\end{gather*}
and in the limit of $R\rightarrow\infty$ we get
\begin{gather}\label{coor2}
\sinh\tau_1\sim\tau_1\sim\frac{y}{R},
\qquad
\sinh\tau_2\sim\tau_2\sim\frac{x}{R},
\end{gather}
where $x$ and $y$ are the Cartesian coordinates in the Euclidean
plane arising from contraction $R\to\infty$.
The orthogonal wave function takes the form~\cite{IPSW2,2,PSW1,PW1}
\begin{gather*}%\label{EQ_solution}
\Psi^{\rm EQ}_{\rho\nu}(\tau_1,\tau_2)=
(\cosh\tau_1)^{-1/2}P^{i\rho}_{-1/2+i\nu}(-\varepsilon\tanh\tau_1)
\exp(i\nu\tau_2),
\end{gather*}
where $\varepsilon=\pm1$.

To perform the contraction we use the formula~\cite{BE1}
\begin{gather}
P^{\mu}_{\nu}(z)=2^\mu \sqrt{\pi}
\big(1-z^2\big)^{-\mu /2}\Biggl
[ \frac{_2F_1\left(-{1\over 2}(\mu +\nu ),{1\over 2}(1-\mu +\nu );
{1\over 2}; z^2\right)}
{\Gamma \left({1\over 2}(1-(\mu +\nu )\right)\Gamma \left(1+{1\over 2}(\nu -\mu )\right)}
\nonumber\\
\phantom{P^{\mu}_{\nu}(z)=}
-2z\frac{_2F_1\left({1\over 2}(1-\mu -\nu ),1+{1\over 2}(\nu -\mu );
{3\over 2}; z^2\right)}
{\Gamma \left({1\over 2}(1+\nu -\mu )\right)\Gamma \left(-{1\over 2}(\nu +\mu )\right)}
\Biggr],\label{L-HYP1}
\end{gather}
and rewrite the Legendre function in terms of the hypergeometric function
\begin{gather*}
P^{i\rho}_{i\nu-1/2}(-\varepsilon\tanh\tau_1)
=\frac{\sqrt{\pi}2^{i\rho}
(\cosh\tau_1)^{i\rho}}
{\Gamma\left(\frac{3}{4}-a\right)
\Gamma\left(\frac{3}{4}-b\right)}
\Biggl\{{_2F_1}\left(\frac{1}{4}-a,
\frac{1}{4}-b;\frac{1}{2};\tanh^2\tau_1\right)
\nonumber\\
\hphantom{P^{i\rho}_{i\nu-1/2}(-\varepsilon\tanh\tau_1)=}{}
+2\varepsilon\tanh\tau_1\frac{\Gamma\left(\frac{3}{4}-a\right)                           %   +- ?
\Gamma\left(\frac{3}{4}-b\right)}
{\Gamma\left(\frac{1}{4}-a\right)
\Gamma\left(\frac{1}{4}-b\right)}
{_2F_1}\left(\frac{3}{4}-a,
\frac{3}{4}-b;\frac{3}{2};\tanh^2\tau_1\right)
\Biggr\},
\end{gather*}
where $a=i(\rho+\nu)/2$, $b=i(\rho-\nu)/2$.
In the contraction limit $R\rightarrow\infty$ we put
\begin{gather*}
\rho\sim kR,\qquad\nu\sim k_1R.
\end{gather*}
Then using also~\eqref{coor2} we have the asymptotic formulae
\begin{gather}
\lim_{R\rightarrow\infty}
{_2F_1}\left(\frac{1}{4}-a,\frac{1}{4}-b;\frac{1}{2};
\tanh^2\tau_1\right)
={_0F_1}\left(\frac{1}{2};-\frac{y^2k_2^2}{4}\right)
=\cos k_2y,
\label{BESSEL-1}
\\
\lim_{R\rightarrow\infty}
{_2F_1}\left(\frac{3}{4}-a,\frac{3}{4}-b;\frac{3}{2};
\tanh^2\tau_1\right)
={_0F_1}\left(\frac{3}{2};-\frac{y^2k_2^2}{4}\right)
=\frac{1}{k_2y}\sin k_2y,
\label{BESSEL-2}
\end{gather}
where $k_1^2+k_2^2=k^2$ give us the contraction limit for
$P^{i\rho}_{i\nu-1/2}(-\varepsilon\tanh\tau_1)$,
\begin{gather*}
P^{i\rho}_{i\nu-1/2}(-\varepsilon\tanh\tau_1)
\approx P^{i k R}_{ik_1R-1/2}
\left(-\varepsilon\frac{y}{R}\right)
\approx\frac{\sqrt{\pi}2^{ik R}\exp(i\varepsilon k_2y)}
{\Gamma\left(\frac{3}{4}-\frac{iR(k+k_1)}{2}\right)
\Gamma\left(\frac{3}{4}-\frac{iR(k-k_1)}{2}\right)},
\end{gather*}
and f\/inally for $\Psi^{\rm EQ}_{\rho\nu}$
\begin{gather*}
\Psi^{\rm EQ}_{\rho\nu}(\tau_1,\tau_2)\approx
\frac{\sqrt{\pi}2^{ik R}}
{\Gamma\left(\frac{3}{4}-\frac{iR(k+k_1)}{2}\right)
\Gamma\left(\frac{3}{4}-\frac{iR(k-k_1)}{2}\right)}
\exp(ik_1x+i\varepsilon k_2y).
\end{gather*}
The last result up to the constant factor coincides with the contraction
limit of equidistant wave function to Cartesian one in Euclidean
space.

\subsection[Horocyclic basis to Cartesian in $E_2$]{Horocyclic basis to Cartesian in $\boldsymbol{E_2}$}

In the notation of the article mentioned horocyclic coordinates on the
hyperboloid can be written $(\bar x\in\mathbb{R},\bar y>0)$
\begin{gather*}
u_0=R {\bar x^2+\bar y^2+1\over 2\bar y},
\qquad
u_1=R {\bar x^2+\bar y^2-1\over 2\bar y},
\qquad
u_2=R {\bar x\over \bar y}.
\end{gather*}
From these relations we see that
\begin{gather*}
\bar x = {u_2\over u_0-u_1},
\qquad
\bar y = {R\over u_0-u_1}.
\end{gather*}
In the limit as $R\rightarrow\infty$ we obtain
\begin{gather*}
\bar x \rightarrow {y\over R},
\qquad
\bar y \rightarrow 1+{x\over R},
\end{gather*}
where $x$ and $y$ are the Cartesian coordinates in the Euclidean plane.
The horicyclic basis functions satisfying the orthonormality condition
\begin{gather*}
R^2\int^{\infty}_{-\infty}d\tilde{x}\int^{\infty}_{0}
\Psi^{\ast {\rm HO}}_{\rho^\prime s^\prime}(\tilde{y},\tilde{x})
\Psi^{\rm HO}_{\rho s}(\tilde{y},\tilde{x})\frac{d\tilde{y}}{\tilde{y}^2}
=
\delta(\rho-\rho^\prime)\delta(s-s^\prime),
\end{gather*}
have the form
\begin{gather*}
\Psi^{\rm HO}_{\rho s}(\bar x,\bar y) =
\sqrt{\frac{\rho \sinh \pi \rho}{ 2 R^2 \pi^3}}
\sqrt{\bar{y}} K_{i\rho }(|s|\bar y) e^{is\bar x},
\end{gather*}
where $K_\nu(x)$ is the Macdonald function~\cite{BE2}.

To ef\/fect the correct contraction we further require that $\rho\rightarrow
kR$, and $s\rightarrow k_2R$ where $k=\sqrt{k^2_1+k^2_2}$.
Consequently we need the asymptotic formula for
\begin{gather*}
K_{ikR}(k_2(R+x))
\end{gather*}
as $R\rightarrow\infty$.
For this we use the asymptotic formula~\cite{BE2}
\begin{gather}\label{K_inu}
K_{i\nu}(x)\sim\frac{\sqrt{2\pi}}{(\nu^2-x^2)^{1/4}}
\exp\left(-\frac{\pi\nu}{2}\right)\sin\left(\frac{\pi}{4}-
\sqrt{\nu^2-x^2}+\nu\cosh^{-1}\frac{\nu}{x}\right),
\end{gather}
valid if $\nu>x\gg 1$ and both $\nu$ and $x$ are large and positive.
If we use this formula for the Macdonald functions and take %ing
into account that
\begin{gather*}
\sqrt{\rho\sinh\pi\rho}
\sim
\sqrt{\frac{k R}{2}}e^{\frac{k\pi R}{2}},
\end{gather*}
then we f\/inally get
\begin{gather*}
\Psi^{\rm HO}_{\rho s}(\bar{y},\bar{x})
\sim
\frac{1}{R\pi}\sqrt{\frac{k}{2k_1}}
\sin(k_1x-M)\exp(i k_2y),
\qquad
M=\frac{\pi}{4}+R\left[k\cosh^{-1}\frac{k}{k_2}-k_1\right].
\end{gather*}
The correct correspondence to give the limiting result in the
Cartesian coordinates $x$ and $y$ can be obtained using the
delta-function contractions
\begin{gather*}
\delta(\rho-\rho^\prime)\delta(s-s^\prime)
\sim\frac{k}{k_1R^2}
\delta(k_1-k_1^\prime)\delta(k_2-k_2^\prime).
\end{gather*}

\subsection{Semi circular parabolic to Cartesian coordinates}

These coordinates are given by the formulae ($\eta,\xi>0$)
\begin{gather*}
u_0=R {\big(\xi^2+\eta^2\big)^2+4\over 8\xi \eta },
\qquad
u_1=R {\big(\xi^2+\eta^2\big)^2-4\over 8\xi \eta },
\qquad
u_2=R {\big(\eta^2-\xi^2\big)\over 2\xi \eta }.
\end{gather*}
In terms of these coordinates we see that
\begin{gather*}
\eta^2 = \frac{\sqrt{R^2+u^2_2}+u_2}{u_0-u_1},
\qquad
\xi^2 = \frac{\sqrt{R^2+u^2_2}-u_2}{u_0-u_1}.
\end{gather*}
In the limit as $R\rightarrow\infty$ we have
\begin{gather*}
\eta^2 = 1 + \frac{x+y}{R},
\qquad
\xi^2 = 1 + \frac{x-y}{R}.
\end{gather*}
We see that in the contraction limit the Cartesian variables
$x$ and $y$ are mixed up.
To have a~correct limit we can introduce the equivalent semi-circular
parabolic system of coordinate connected with previous one by
the rotation about axis $u_0$ through the angle $\pi/4$,  %:                          % rotation about ? the axis?
\begin{gather*}%\label{sys_scp_xi_eta_rot}
u_0 = R\frac {\left(\eta^2 + \xi^2\right)^2 + 4}{8 \xi \eta},
\qquad
u_1 = R \frac{\sqrt{2}}{2}\left(\frac {{\eta}^{2}-{\xi}^{2}}{2\xi\eta}
+ \frac {\left(\eta^2 + \xi^2\right)^2 - 4}{8 \xi \eta} \right),
\\
u_2 = R \frac{\sqrt{2}}{2}\left( \frac {{\eta}^{2}-{\xi}^{2}}{2\xi\eta}
-\frac {\left(\eta^2 + \xi^2\right)^2 - 4}{8 \xi \eta} \right).
\end{gather*}
In terms of a~new coordinates we have that
\begin{gather*}
\eta^2=\frac{\sqrt{2R^2+(u_2-u_1)^2}-(u_2-u_1)}
{\sqrt{2}u_0-(u_2+u_1)},
\qquad
\xi^2=\frac{\sqrt{2R^2+(u_2-u_1)^2}+(u_2-u_1)}
{\sqrt{2}u_0-(u_2+u_1)},
\end{gather*}
and in the contraction limit $R\to\infty$ we obtain
\begin{gather*}%\label{contr_SCP}
\eta^2\to1+\sqrt{2}\frac{x}{R},\qquad
\xi^2\to1+\sqrt{2}\frac{y}{R}.
\end{gather*}
The suitable set of basis functions are~\cite{GROSCHE, 2}
\begin{gather*}
\Psi^{\rm SCP}_{\rho s}(\xi, \eta ) = \sqrt{\xi\eta}
J_{i \rho }\left(\sqrt{s} \xi\right) K_{i \rho }\left(\sqrt{s} \eta\right),
\end{gather*}
for $s>0$ and
\begin{gather*}
\Psi^{SCP}_{\rho s}(\xi, \eta ) = \sqrt{\xi\eta}
K_{i \rho }\left(\sqrt{-s} \xi\right) J_{i \rho }\left(\sqrt{-s} \eta\right),
\end{gather*}
for $s<0$.
The correct limit is then obtained by choosing
$s=R^2(k_2^2-k_1^2)$ and $\rho=k R$, where $k_1^2+k_2^2=k^2$.
To f\/ind the contraction limit of the basis function let us use the
asymptotic relation for the Bessel function of pure imaginary index~\cite{Watson:1944}
\begin{gather*}
2\pi J_{ip}(z)\sim\frac{\sqrt{2\pi}}{(p^2+z^2)^{1/4}}
\exp\left(i\sqrt{p^2+z^2}-i p\sinh^{-1}\frac{p}{z}-
i\frac{\pi}{4}\right)\exp\left(\frac{p\pi}{2}\right),
\end{gather*}
and for the MacDonald function the formula~\eqref{K_inu}.
Then
\begin{gather*}
J_{i\rho}\left(\sqrt{s}\xi\right)\sim J_{i k R}\left(R\sqrt{k_2^2-k_1^2}\sqrt{1+\sqrt{2}
\frac{y}{R}}\right)
\nonumber\\
\phantom{J_{i\rho}\left(\sqrt{s}\xi\right)}
\sim\frac{e^{\frac{\pi}{2}kR}}{2^{\frac{1}{4}}\sqrt{2\pi k_2R}}
\exp\left(i k_2y+i R\left[\sqrt{2}k_2-k\sinh^{-1}\frac{k}{\sqrt{k_2^2-k_1^2}}
\right]-i\frac{\pi}{4}\right)
\end{gather*}
and
\begin{gather*}
K_{i\rho}(\sqrt{s}\eta)\sim K_{ikR}\left(R\sqrt{k_2^2-k_1^2}\sqrt{1+\sqrt{2}\frac{x}{R}}\right)
\nonumber\\
\phantom{K_{i\rho}(\sqrt{s}\eta)}
\sim
\frac{2^\frac{1}{4}\sqrt{\pi}e^{-\frac{\pi}{2}kR}}{\sqrt{R k_1}}
\sin\left(\frac{\pi}{4}-x k_1+R\left[\sqrt{2}k\cosh^{-1}\frac{k}{\sqrt{k_2^2-k_1^2}}-k_1\right]\right).
\end{gather*}
Using the last asymptotic formulae it is easily to get for a~large $R$
\begin{gather*}%\label{contr_sol_SCP}
\Psi^{\rm SCP}_{\rho s}(\xi,\eta)
\sim
\frac{-1}{R\sqrt{2k_1k_2}}
\exp\left(i k_2y+i\delta_1-i\frac{\pi}{4}\right)
\sin\left(k_1x-\frac{\pi}{4}+\delta_2\right),
\end{gather*}
where
\begin{gather*}
\delta_1+\delta_2
=\sqrt{2}R(k_1+k_2)-Rk \sinh^{-1}\left(\frac{k_2+k_1}{k_2-k_1}
\right),
\\
\delta_1-\delta_2
=\sqrt{2}R(k_2-k_1)-Rk \sinh^{-1}\left(\frac{k_2-k_1}{k_2+k_1}
\right).
\end{gather*}
These expressions are arrived at under the assumption that $k^2_2>k^2_1$.
In case of $k_2^2<k_1^2$ we can make the interchanges $x$ with $y$,
and $k_1$ with $k_2$.

\subsection{Elliptic parabolic basis to parabolic}

Elliptic parabolic basis contracts into a parabolic one on~$E_2$.
In these coordinates the points on the hyperbola are given by
$[\theta\in(-\pi/2,\pi/2),a\ge0]$,
\begin{gather*}
u_0= R {\cosh^2a+ \cos^2\theta \over 2\cosh a~\cos\theta },
\qquad
u_1= R {\sinh^2a- \sin^2\theta \over 2\cosh a~\cos\theta },
\qquad
u_2= R \tanh a~\tan\theta.
\end{gather*}
From these relations we see that
\begin{gather*}
\cos^2\theta = \frac{u_0- \sqrt{u^2_0-R^2}}{u_0-u_1},
\qquad
\cosh^2 a~= \frac{u_0 + \sqrt{u^2_0-R^2}}{u_0-u_1}.
\end{gather*}
In the limit as $R\rightarrow\infty$ we obtain
\begin{gather*}
\cos^2\theta \rightarrow 1- {\eta^2\over R},
\qquad
\cosh^2a \rightarrow 1+ {\xi^2\over R},
\end{gather*}
where the parabolic coordinates $(\xi,\eta)$ are given by
\begin{gather*}
x={1\over 2}\big(\xi^2-\eta^2\big),
\qquad
y=\xi \eta,
\qquad
\xi, \eta > 0.
\end{gather*}
The elliptic parabolic wave functions on the hyperboloid have the form
\cite{GROSCHE}
\begin{gather*}
\Psi^{\rm EP}_{\rho s}(a, \theta ) = \sqrt{\cos\theta}
P^{i\rho }_{is-1/2}(\sin\theta)P^{is}_{i\rho -1/2}(\tanh a).
\end{gather*}
To ef\/fect the contraction we take $s\rightarrow\kappa R$ and
$\rho\rightarrow k R$ and
\begin{gather*}
\tanh a~\sim \frac{\xi}{\sqrt{R}},
\qquad
\sin\theta \sim \frac{\eta}{\sqrt{R}}.
\end{gather*}
The separation equation
\begin{gather*}%\label{ch}
\left(\frac{d^2}{d a^2} + s^2 - \frac{\rho^2 + 1/4}
{\cosh^2 a}\right) F(a) = 0,
\end{gather*}
becomes                                                               % what becomes?
\begin{gather*}
\left(\frac{d^2}{d \xi^2} + \lambda + k^2 \xi^2 \right) F(\xi) = 0,
\end{gather*}
where $\lambda$ is now the parabolic separation constant and
we impose the condition
\begin{gather*}
\lambda = R \big(\kappa^2-k^2\big),
\end{gather*}
or
\begin{gather*}
\kappa=k+{\lambda\over2kR}+O\big(R^{-2}\big).
\end{gather*}
This requires taking %us to take
the limit of $a$ and $\theta$    % in the dependent part ?
the dependent part of the eigenfunctions $P^{is}_{i\rho-1/2}(\tanh a)$
and $P^{i\rho}_{is-1/2}(\sin\theta)$.
The limit can be established
from the known representation of the Legendre function~\eqref{L-HYP1}.
This results in the asymptotic formula
\begin{gather*}
P^{ikR+{i\lambda \over 2k}}_{ikR-{1\over 2}}\left(\frac{\xi}{\sqrt{R}}\right)
\rightarrow \frac{2^{ikR+{1\over 4}+{i\lambda \over 4k}}}
{\Gamma\left({3\over 4}- {i\lambda \over 4k}-ikR\right)}
D_{{1\over 2}\left(i{\lambda \over k} -1 \right)}\big(\sqrt{-2ik} \xi\big),
\end{gather*}
where $D_\nu(z)$ is a~parabolic cylinder function~\cite{BE1}
\begin{gather*}
D_\nu(z)=2^{\nu/2}\sqrt{\pi}e^{-\frac{z^2}{4}}
\left[
\frac{1}{\Gamma\left(\frac{1-\nu}{2}\right)}
{_{1}F_{1}}\left(-\frac{\nu}{2};\frac{1}{2};\frac{z^2}{2}\right)-
\frac{z\sqrt{2}}{\Gamma\left(-\frac{\nu}{2}\right)}
{_{1}F_{1}}\left(\frac{1-\nu}{2};\frac{3}{2};\frac{z^2}{2}\right)
\right].
\end{gather*}
If we look at the theta-dependent part of the eigenfunctions and
the corresponding limit taking $\sin\theta={\eta\over\surd R}$
we then obtain the limit
\begin{gather*}
P^{ikR}_{ikR+{i\lambda \over 2k} -{1\over 2}}\left(\frac{\eta}{\sqrt{R}}\right)
\rightarrow \frac{2^{ikR+{1\over 4}+{i\lambda \over 4k}}}
{\Gamma \left({3\over 4}- {i\lambda \over 4k} -ikR\right)}
D_{-{1\over 2}\left(i{\lambda \over k} +1\right)}\big(\sqrt{-2ik} \eta \big),
\end{gather*}
and f\/inally
\begin{gather*}
\Psi^{\rm EP}_{\rho s}(a, \theta ) \sim \frac{
2^{2ikR+{1\over 2}+{i\lambda \over 2k}}}
{\left[\Gamma \left({3\over 4}-{i\lambda \over 4k} -ikR\right)\right]^2}
D_{-{1\over 2}\big(i{\lambda \over k} +1\big)}\big(\sqrt{-2ik} \eta \big)
D_{{1\over 2}\big(i{\lambda \over k} -1 \big)}\big(\sqrt{-2ik} \xi\big).
\end{gather*}
From these expressions we see that we do indeed obtain the correct
asymptotic limit.

\subsection{Hyperbolic parabolic to a~Cartesian basis}

Hyperbolic parabolic basis contracts into a~Cartesian one on~$E_2.$
In these coordinates the points on the hyperbola are given by
[$\theta\in(0,\pi)$, $b>0$],
\begin{gather*}
u_0=R {\cosh^2b + \cos^2\theta \over 2\sinh a~\sin\theta },
\qquad
u_1=R {\sinh^2b - \sin^2\theta \over 2\sinh a~\sin\theta },
\qquad
u_3=R \cot\theta \coth b.
\end{gather*}
From these relations we see that
\begin{gather*}
\cos^2\theta = \frac{u_0- \sqrt{u^2_1+R^2}}{u_0-u_1},
\qquad
\cosh^2b = \frac{u_0 + \sqrt{u^2_1+R^2}}{u_0-u_1}.
\end{gather*}
In the limit as $R\rightarrow\infty$ we can choose
\begin{gather*}
\cos^2\theta \rightarrow \frac{y^2}{2R^2},
\qquad
\cosh^2 b \rightarrow 2\left(1+{x\over R}\right).
\end{gather*}
The hyperbolic parabolic basis function on hyperbolid can be chosen
in the form~\cite{GROSCHE}
\begin{gather}\label{2SH_HP_sol}
\Psi^{\rm HP}(b,\theta ) = (\sinh b \sin\theta )^{1/2}P^{i\rho}_{is-1/2}(\cosh b ) P^{i\rho}_{is-1/2}(\cos\theta ).
\end{gather}
To proceed further with this limit we take $\rho^2\sim k^2R^2$ and
$s^2\sim\big(k_1^2-k_2^2\big)R^2$ (the case of $s^2<0$, or
$k_1^2<k_2^2$, corresponds to the discrete spectrum of constant $s$,
and we do not consider this case here) where $k_1^2+k_2^2=k^2$, then
using the relation between Legendre function and hypergeometric
functions~\eqref{L-HYP1} and formulae \eqref{BESSEL-1}, and \eqref{BESSEL-2},
we obtain for $\theta$ depending part of basis function
\begin{gather*}
\sqrt{\sin\theta}P^{i\rho}_{is-1/2}(\cos\theta)
\sim
P^{i k R}_{i\sqrt{k_1^2 - k_2^2} R - 1/2} \left(\frac{y}{\sqrt{2} R}\right)
\\
\phantom{\sqrt{\sin\theta}P^{i\rho}_{is-1/2}(\cos\theta)}
\sim
\frac{2^{ik R}\sqrt{\pi}\exp(i k_2 y)}
{\Gamma\left(\frac{3}{4} - \frac{i R}{2}\left(k + \sqrt{k_1^2 - k_2^2}\right) \right)
\Gamma\left(\frac{3}{4} - \frac{i R}{2}\left(k - \sqrt{k_1^2 - k_2^2}\right) \right)}.
\end{gather*}
For the limit of the $b$ dependent part of the eigenfunctions we must
proceed dif\/ferently.
In fact we need to calculate the limit of
\begin{gather*}
P^{i k R}_{i\sqrt{k_1^2 - k_2^2} R - 1/2}\left(\sqrt{2\left(1+{x\over R}\right)}\right)
\end{gather*}
as $R\rightarrow\infty$.
We know that the leading terms of this expansion
have the form
\begin{gather*}
A\exp(i k_1x)+B\exp(-i k_1x),
\end{gather*}
and we now make use of this fact.
By this we mean that
\begin{gather*}
\lim_{R\rightarrow\infty}
P^{i k R}_{i\sqrt{k_1^2-k_2^2}R-1/2}\left(\sqrt{2\left(1+{x\over R}\right)}\right)
=A\exp(i k_1x)+B\exp(-i k_1x),
\end{gather*}
where the constants $A$ and $B$ depend on $R$.
It remains to determine $A$ and $B$.
To do this let us
consider $x=0$.
We then need to determine the following limit
\begin{gather}\label{ApB}
\lim_{R\to \infty} P_{ -\frac{1}{2} + i R \sqrt{ k_1^2 - k_2^2}}^{i k R}\big( \sqrt{2}\big) = A + B.
\end{gather}
%This can be done using the method of stationary phase.
%This can be seen as follows.
From the integral representation formula
\cite{BE1}
\begin{gather}
\frac{\Gamma(-\nu -\mu)\Gamma(1+\nu -\mu)}{\Gamma (1/2- \mu)}
{\sqrt\frac{\pi}{2}}
P^\mu _\nu (z)
\nonumber\\
\qquad=\big(z^2-1\big)^{-\mu /2}\int^\infty_0 \left(z+\cosh t\right)^{\mu -1/2}
\cosh\left(\left[\nu + {1/2}\right]t\right){\mathrm d}t,\label{LEGANDRE1}
\end{gather}
the above limit requires us to calculate as $R\rightarrow\infty$                    % requires us
\begin{gather*}
\int^\infty_0\big(\sqrt{2}+ \cosh t\big)^{ikR-1/2}
\cos \left(\left[ R \sqrt{ k_1^2 - k_2^2}\right] t\right){\mathrm d}t.
\end{gather*}
It can be done using the method of stationary phase~\cite{Olver:74}.
We obtain
\begin{gather}
P_{- \frac{1}{2} + iR \sqrt{k_1^2 - k_2^2}}^{i k R} \big( \sqrt{2}\big) \sim
\frac{ 2^{ -\frac{5}{4} + \frac{iR}{2}\left( \sqrt{k_1^2 - k_2^2} - k\right)}
\Gamma \left(\frac{1}{2}-i k R \right) } {\Gamma \left[ \frac{1}{2} - iR
\left(\sqrt{k_1^2 - k_2^2} + k\right) \right]\Gamma \left[ \frac{1}{2} +
i R\left(\sqrt{k_1^2 - k_2^2}-k\right)\right]}
\nonumber\\
\hphantom{P_{- \frac{1}{2} + iR \sqrt{k_1^2 - k_2^2}}^{i k R} \big( \sqrt{2}\big) \sim}{}
\times
\left(\frac{i}{Rk_1}\right)^{1/2} \left( \frac{k_1 - \sqrt{k_1^2 - k_2^2}}
{k + \sqrt{k_1^2 - k_2^2}}\right)^{i R \sqrt{k_1^2 - k_2^2}}
\left(\frac{k}{k-k_1} \right)^{i k R}.\label{KK}
\end{gather}
By considering the expression for the derivatives of the Legendre function~\eqref{LEGANDRE1}
at $x=0$, we derive the expression
\begin{gather*}
\left.
\frac{d}{dx} P_\nu^\mu(z)\right |_{x=0} \sim - i k_1
P_{-\frac{1}{2} + i R\sqrt{k_1^2 - k_2^2}}^{i k R} \big( \sqrt{2}\big) \sim i k_1 (A-B),
\end{gather*}
then
\begin{gather*}
P_{-\frac{1}{2}+i R\sqrt{k_1^2-k_2^2}}^{i k R}\big(\sqrt{2}\big)\sim-A+B.
\end{gather*}
Comparing the above relation with~\eqref{ApB}, we obtain that $A=0$ and $B$ is equal to \eqref{KK},
that is
\begin{gather*}
P_{is-\frac{1}{2}}^{i\rho}(\cosh b)\to B e^{-i x k_1}.
\end{gather*}
Finally, solution~\eqref{2SH_HP_sol} contracts as follows
\begin{gather*}
\Psi_{\rho s}(b,\theta)\sim\frac{2^{ik R}\sqrt{\pi}B}
{\Gamma\left(\frac{3}{4}-i R\frac{k+\sqrt{k_1^2-k_2^2}}{2}\right)
\Gamma\left(\frac{3}{4}-i R\frac{k-\sqrt{k_1^2-k_2^2}}{2}\right)}
\exp(i k_2y-i k_1x).
\end{gather*}

\section{Conclusion}

In this note we have constructed the contraction limit $R\to\infty$
for the unnormalized wave functions which are the solution of the Helmholtz equation
on the two dimensional two sheeted hyperboloid in four coordinates systems,
namely, horocyclic, semi circular parabolic, elliptic parabolic and hyperbolic
parabolic.
Of course the complete analysis of the contraction problem must
include also the solutions of Helmholtz equation in the three additional systems
of coordinates as elliptic, hyperbolic and semi hyperbolic.
We will study this in the near future.

We have not presented limits associated with nonsubgroup coordinates.
The extension of these ideas to problems in higher dimensions is natural and will
be presented in subsequent work.

\subsection*{Acknowledgment}

The work of G.S.P.\ was partially supported under the Armenian-Belorussian grant
11RB-010.
G.S.P.\ and A.Ya.\ are thankful to the PRO-SNI (UdeG, Mexico).

\pdfbookmark[1]{References}{ref}

\LastPageEnding

\end{document}